\documentclass[
	onecolumn,
	10pt,
	]{article}
\usepackage{extsizes}
\usepackage[super,sort&compress,comma]{natbib} 
\usepackage[version=3]{mhchem}
\usepackage[left=2.5cm, right=2.5cm, top=1.785cm, bottom=2.0cm]{geometry}
\usepackage{balance}
\usepackage{times,mathptmx}
\usepackage{sectsty}
\usepackage{graphicx} 
\usepackage{lastpage}
\usepackage[format=plain,justification=raggedright,singlelinecheck=false,font={stretch=1.125,small,sf},labelfont=bf,labelsep=space]{caption}
\usepackage{float}
\usepackage{fancyhdr}
\usepackage{fnpos}
\usepackage[english]{babel}
\usepackage{array}
\usepackage{droidsans}
\usepackage{charter}
\usepackage[T1]{fontenc}
\usepackage[usenames,dvipsnames]{xcolor}
\usepackage{setspace}
\usepackage[compact]{titlesec}
\usepackage{xspace}
\usepackage{multirow}
\usepackage{placeins}
\makeatletter
\def\hlinewd#1{%
  \noalign{\ifnum0=`}\fi\hrule \@height #1 \futurelet
   \reserved@a\@xhline}
\makeatother

\newcommand{\LiZn}{LiZn$_2$Mo$_3$O$_8$\xspace}

\newcommand{\MoO}{Mo$_3$O$_{13}$\xspace}

\newcommand{\NbC}{Nb$_3$Cl$_8$\xspace}
\newcommand{\NbCC}{Nb$_3$Cl$_{13}$\xspace}
\newcommand{\Shalf}{\textit{S$_{eff}$}~=~1/2\xspace}
\newcommand{\Teq}{\textit{T}~=~}
\newcommand{\Tapp}{\textit{T}~$\approx$~}
\newcommand{\K}{\,K\xspace}
\newcommand{\Ptm}{\textit{P}$\bar{3}$\textit{m}$1$\xspace}
\newcommand{\Ctm}{\textit{C}2/\textit{m}\xspace}
\newcommand{\degC}{$^\textrm{o}$C\xspace}
\newcommand{\muoH}{$\mu_{o}H$~=~}
\newcommand{\GaNbS}{GaNb$_4$S$_8$\xspace}

\newcommand{\BaCu}{Ba$_3$CuSb$_2$O$_9$\xspace}

\newcommand{\BaCo}{Ba$_3$CoSb$_2$O$_9$\xspace}

\definecolor{cream}{RGB}{222,217,201}

\begin{document}

\makeatletter 
\newlength{\figrulesep} 
\setlength{\figrulesep}{0.5\textfloatsep} 

\newcommand{\topfigrule}{\vspace*{-1pt}%
\noindent{\color{cream}\rule[-\figrulesep]{\columnwidth}{1.5pt}} }

\newcommand{\botfigrule}{\vspace*{-2pt}%
\noindent{\color{cream}\rule[\figrulesep]{\columnwidth}{1.5pt}} }

\newcommand{\dblfigrule}{\vspace*{-1pt}%
\noindent{\color{cream}\rule[-\figrulesep]{\textwidth}{1.5pt}} }

\makeatother


\vspace{3cm}

\begin{center}
\noindent\LARGE{\textbf{Rearrangement of Van-der-Waals Stacking and Formation of a Singlet State at \Teq90\K in a Cluster Magnet}} \\

\vspace{0.6cm} 

\noindent\large{John P. Sheckelton,\textit{$^{a,b}$} Kemp W. Plumb,\textit{$^{b}$} Benjamin A. Trump,\textit{$^{a,b}$} Collin L. Broholm,\textit{$^{b,c,d}$} and Tyrel M. McQueen\textit{$^{a,b,c,*}$}} \\
\end{center}

\begin{changemargin}{0.6cm}{0.6cm}
\noindent\normalsize{Insulating \NbC is a layered chloride consisting of two-dimensional triangular layers of \Shalf \NbCC clusters at room temperature. Magnetic susceptibility measurement show a sharp, hysteretic drop to a temperature independent value below \Teq90\K. Specific heat measurements show that the transition is first order, with $\Delta S \approx 5\ \mathrm{J \cdot K^{-1} \cdot mol\ \mathit{f.u.}^{-1}}$, and a low temperature \textit{T}-linear contribution originating from defect spins. Neutron and \mbox{X-ray} diffraction show a lowering of symmetry from trigonal \Ptm to monoclinic \Ctm symmetry, with a change in layer stacking from \mbox{-AB-AB-} to \mbox{-AB$^\prime$-BC$^\prime$-CA$^\prime$-} and no observed magnetic order. This lowering of symmetry and rearrangement of successive layers evades geometric magnetic frustration to form a singlet ground state. It is the lowest temperature at which a change in stacking sequence is known to occur in a Van-der-Waals solid, occurs in the absence of orbital degeneracies, and suggests that designer 2-D heterostructures may be able to undergo similar phase transitions.}\\
\end{changemargin}



\footnotetext{\textit{$^{a}$~Department of Chemistry, The Johns Hopkins University, Baltimore, MD 21218, USA.}}
\footnotetext{\textit{$^{b}$~Institute for Quantum Matter and Department of Physics and Astronomy, The Johns Hopkins University, Baltimore, MD 21218, USA.}}
\footnotetext{\textit{$^{c}$~Department of Materials Science and Engineering, The Johns Hopkins University, Baltimore, Maryland 21218, USA.}}
\footnotetext{\textit{$^{d}$~NIST Center for Neutron Research, National Institute of Standards and Technology, Gaithersberg, MD 20899, USA.}}
\footnotetext{$^*$~E-mail:\,\textit{mcqueen@jhu.edu}}


\section{Introduction}

Emergent phenomena among strongly interacting atoms or electrons, such as superconductivity\cite{rotter_superconductivity_2008,neto_ubiquitous_2014,nagamatsu_superconductivity_2001}, charge density waves\cite{moncton_neutron_1977}, topological insulators\cite{chen_experimental_2009,hsieh_observation_2009}, Kondo insulators\cite{iga_single_1998,mason_spin_1992}, and heavy fermions\cite{palstra_superconducting_1985}, are at the forefront of contemporary materials research. Geometrically frustrated magnets are a particularly illustrative class of strongly interacting systems where a large degeneracy of electronic states exist within a small energy regime compared to the magnetic interaction strength. Since the electronic degeneracy arises from lattice symmetry, geometrical frustration can destabilize the lattice. Here we show that a geometrically frustrated antiferromagnet built from small transition metal clusters\cite{sheckelton_possible_2012,sheckelton_local_2014,sheckelton_electronic_2014,flint_emergent_2013,chen_spin_2014,mourigal_molecular_2014} can succumb to a symmetry-lowering distortion to evade a degenerate magnetic ground state, even in the absence of orbital degeneracies. This phase change also involves a change in stacking sequence between successive charge-neutral Van-der-Waals (VdW) bonded layers.

Specifically, we report the discovery of a paramagnetic and trigonal to singlet and monoclinic phase transition in the cluster-based magnet \NbC, despite each \NbCC cluster harboring a singly occupied, non-degenerate highest occupied molecular orbital (HOMO) and an approximately 1\,eV gap to degenerate lowest unoccupied molecular orbital (LUMO) states. As for \MoO clusters in \LiZn\cite{sheckelton_possible_2012}, a formal electron count yields one \Shalf magnetic electron per \NbCC cluster, which are arranged on a two-dimensional triangular lattice. \NbC and various stacking variations\cite{kennedy_chemical_1991,kennedy_experimental_1996,miller_solid_1995,schnering_kristallstruktur_1961,cotton_discrete_1988} have been previously studied. The \mbox{$\alpha$-\NbC} polymorph, with \mbox{-AB-AB-} stacking, is known to undergo a hysteretic magnetic transition\cite{kennedy_-nb3cl8_1992} with a change in the magnetic signal at temperatures below \Tapp90\K. Here, this transition is studied \textit{via} detailed structural and physical property investigations of both powder and single crystals, revealing a dramatic trigonal to monoclinic phase transition at \Teq90\K that quenches the magnetic response but without magnetic order, \textit{i.e.} a singlet state. The relief of geometric frustration \textit{via} orbital ordering and formation of magnetic order or spin singlets, is well-known in compounds containing first-order Jahn-Teller (FOJT) active ions or clusters, such as NaTiO$_{2}$\cite{clarke_synthesis_1998,ezhov_orbital_1998}, NaVO$_{2}$\cite{mcqueen_successive_2008}, or \GaNbS\cite{jakob_structural_2007}. A similar spin-Peierls distortion in 1D \Shalf systems, such as CuGeO$_3$\cite{hase_observation_1993}, NaTiSi$_2$O$_6$\cite{isobe_novel_2002}, and the titanium oxychlorides/oxybromides\cite{shaz_spin-peierls_2005,van_smaalen_incommensurate_2005}, can also be explained in terms of orbital ordering \cite{seidel_chains_2003}.

The phase transition to evade geometric magnetic frustration in \NbC appears to proceed \textit{via} a different route. The structural phase transition breaks the $C_{3v}$ symmetry of the \NbCC clusters in a manner consistent with a second-order Jahn-Teller (SOJT) distortion, but with a dramatic change in \NbC layer stacking from \mbox{-AB-AB-} to \mbox{-AB$^\prime$-BC$^\prime$-CA$^\prime$-}. That such a change in stacking of a VdW material can occur near liquid nitrogen temperature is remarkable, but can be thought of as being driven by a buckling of interfacial Cl-atom layers due to an inter-layer electronic interaction and SOJT distortion combined with singlet formation. Our results demonstrate the importance of considering multi-site effects and states far from the HOMO in magnetically frustrated materials, and that changes in VdW stacking sequence are possible well below room temperature.

\begin{figure}[t!]
	\centering
		\includegraphics[width=8cm]{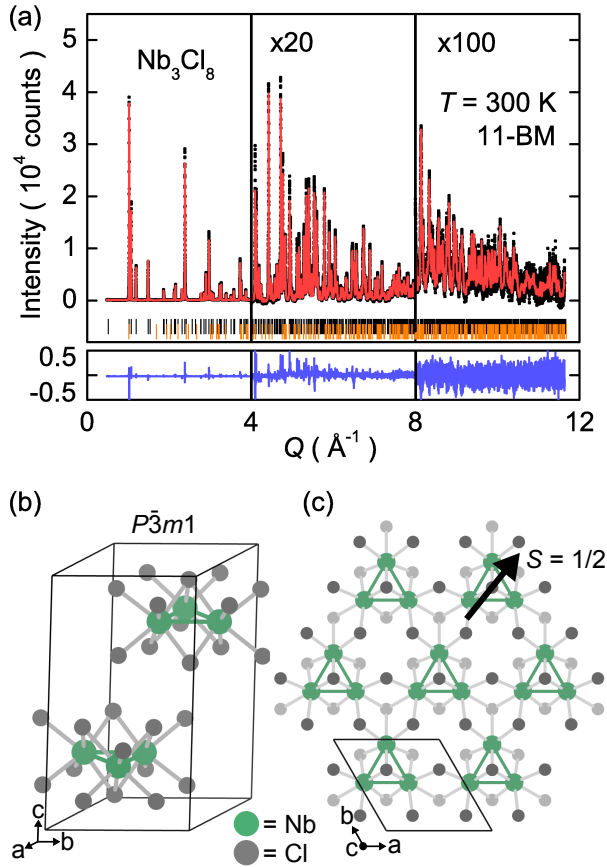}
			\caption{
			(a) Rietveld analysis of synchrotron \mbox{X-ray} diffraction data at \Teq300\K. Black dots are data, red line is the calculated fit, blue line is the difference, tick marks are Bragg reflections for \NbC (black) and a minor impurity, NbOCl$_2$ (orange, 0.3(2) wt\%). The higher $Q$ data are multiplied $\times\!20$ ($4 \geq Q \geq 8$) and $\times\!100$ ($8 \geq Q \geq 12$). The resulting unit cell of the Rietveld analysis is shown in (b) and (c). There are two \NbCC clusters per unit cell with each cluster composed of three edge-sharing NbCl$_6$ octahedra meeting at an apical Cl atom atop the Nb$_3$ triangle. The resulting \NbC layers are inequivalent as can be seen by the alternating direction of the capping Cl atom and thus form an \mbox{-AB-AB-} stacking sequence. A single \textit{ab}-plane of \NbC is shown in (c), forming a triangular lattice of \Shalf triangular clusters.
			}
		\label{Fig:HT_XRD}
\end{figure}

\section{Methods}

Single crystalline \NbC was synthesized by self vapor transport in an evacuated quartz tube charged with a stoichiometric amount of twice re-sublimed NbCl$_5$ (Strem, 99.99\%) and powder Nb metal, used as is (Alfa, 99.99\%). A temperature gradient of \Teq825\degC to \Teq835\degC was maintained over the reaction vessel for 12 to 14 days before cooling to room temperature. All samples were handled in a glovebox using standard air-free techniques. High-resolution synchrotron \mbox{X-ray} diffraction measurements (SXRD) were taken on powder samples of ground single crystals at temperatures from \Teq300\K to \Tapp90\K using the powder diffractometer at beamline 11-BM equipped with a liquid nitrogen cryostream at the Advanced Photon Source (APS). Triple-axis Neutron diffraction measurements of co-aligned \NbC single crystals were acquired from \Teq300\K to \Teq1.6\K on the SPINS spectrometer at the NIST Center for Neutron Research (NCNR). SPINS was operated with a fixed energy of 5 meV, flat analyzer, and a $^{58}$Ni guide - 80' - 40' - open collimation sequence. Measurements were performed with the \NbC single crystal array oriented in both the ($hhl$) or ($h0l$) scattering planes of the high temperature structure. Low-temperature powder \mbox{X-ray} diffraction (PXRD) patterns were acquired from \Teq300\K to \Teq12\K using a Bruker D8 Advance powder diffractometer with Cu$K\alpha$ radiation ($\lambda=1.5424$\,\AA), a scintillator point detector with 0.6\,mm slits, and an Oxford Cryosystems PheniX low-temperature closed cycle cryostat. Powder neutron diffraction at \Teq300\K and \Teq10\K was performed at the POWGEN diffractometer, Spallation Neutron Source, Oak Ridge National Laboratory (ORNL). Scans were measured for 4 hours each on a 0.2\,g sample of crushed \NbC single crystals. Rietveld refinements to synchrotron and in-house \mbox{X-ray} diffraction data were performed using the General Structure Analysis System (GSAS)\cite{schmidt_general_1993} and the commercial Bruker Topas software suite. Refinement to the low-temperature powder neutron diffraction data was performed using the program FAULTS\cite{casas_faults_2006} software package to account for random HT-phase stacking faults \mbox{(-AB-AB-)} in the LT phase. Angle dependent magnetic susceptibility measurements on single crystalline \NbC were performed on a Quantum Design Physical Properties Measurement System (PPMS) from \Teq300\K to \Teq2\K under an applied field of \muoH5\,T, after first measuring the sample holder temperature dependent background, which was subsequently subtracted from the data. Specific heat capacity measurements were taken on a PPMS from \Teq300\K to \Teq2\K using the semi-adiabatic pulse technique and dual slope analysis method. For sensitivity to latent heat, measurements around \Tapp90\K were performed using a single large heat pulse from \Teq85\K to \Teq110\K, and analyzed \textit{via} the multi-point single slope method\cite{QDPPMS}. Resistivity measurements were taken using the four-probe method in the PPMS by attaching Pt wire to single crystalline \NbC using DuPont 4922N silver composition paint. The sample was measured from \Teq300\K to \Tapp275\K where the resistivity exceeded the instrument threshold due to a voltmeter impedance of $\approx 16$\,M$\Omega$. Band structure calculations were performed on the high- and low-temperature phases of \NbC. Convergence was achieved with a $8\!\times\!8\!\times\!4$ and $4\!\times\!7\!\times\!4$ $k$-point mesh for the high- and low-temperature phases respectively, using the ELK all-electron full-potential linearized augmented-plane wave (FP-LAPW) code using the Perdew-Wang/Ceperley-Alder LSDA functional\cite{ELK}.

\section{Results}

Results of Rietveld refinements to powder SXRD data at \Teq300\K of the high-temperature (HT) \Ptm phase are shown in Fig.~\ref{Fig:HT_XRD}. The HT phase unit cell [Fig.~\ref{Fig:HT_XRD}\,(b)] consists of two \NbCC clusters per unit cell in a \mbox{-AB-AB-} bilayer stacking order where A and B refer to independent \NbC layers and AB constitutes a single bilayer, since adjacent \NbC layers are inequivalent due to alternating directions of the Nb$_3$ capping Cl atoms. No atoms occupy the two inequivalent spaces between \NbC layers, leaving only Van-der-Waals (VdW) interactions between interfacial Cl layers as the net attractive force. Rietveld refinement to powder synchrotron diffraction data [Fig.~\ref{Fig:HT_XRD}\,(a)] show low-\textit{Q} diffraction peaks that are asymmetric and not fit well (broad calculated peaks) due to a combination of instrumental low angle peak asymmetry and significant stacking faults. The excellent overall quality of the fit is highlighted by the zoomed regions. To assess whether or not a more subtle distortion is present at high temperatures, fits to space groups with symmetry elements systematically removed (\textit{P}$\bar{3}$, \textit{P}3\textit{m}, \textit{P}3, and \Ctm) do not improve the fit, either statistically or qualitatively. The low temperature synchrotron data near \Tapp90\K show no indication of a symmetry-lowering structural distortion (see ESI$^\dag$). These results from the HT phase are in agreement with the structure previously inferred from single crystal \mbox{X-ray} diffraction\cite{strobele_struktur_2001,schnering_kristallstruktur_1961}.

\begin{figure}[t!]
	\centering
		\includegraphics[width=8cm]{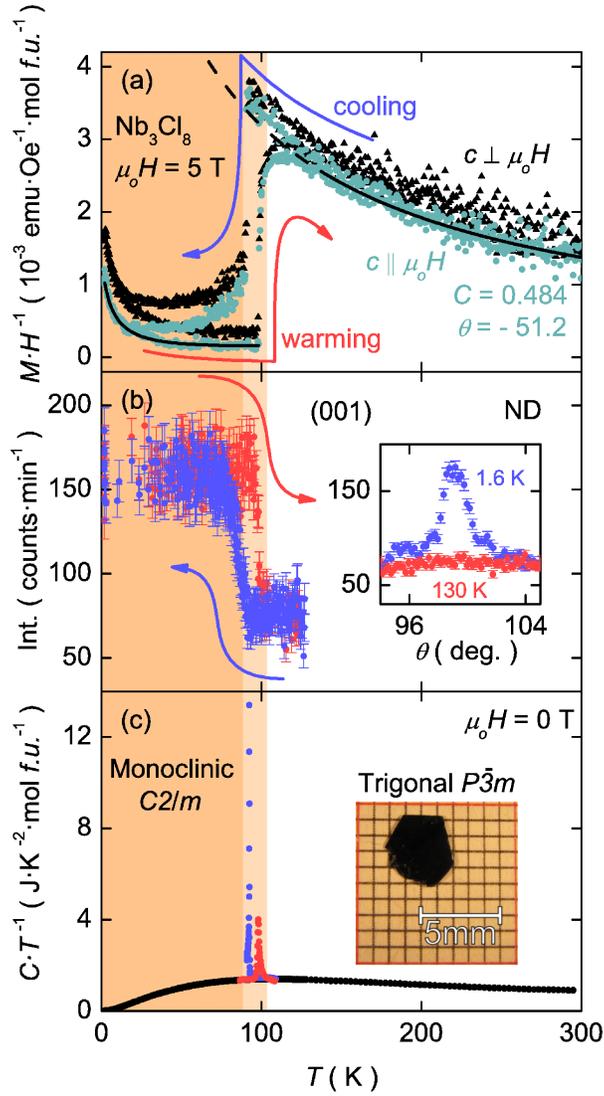}
			\caption{
			(a) \NbC susceptibility with applied field parallel ($\parallel$) and perpendicular ($\perp$) to the crystallographic \textit{c}-direction. Curie-Weiss fits to $c\,\parallel\,\mu_{o}H$ data is shown as a black line, for both high- and low-temperature regions. (b) Neutron diffraction (ND) intensity of the (001) reflection of \NbC, which grows upon entering the \Ctm phase. The inset shows the (001) peak at \Teq1.6\K and its absence at \Teq130\K. (c) Heat capacity over temperature vs. temperature for \NbC. Peaks around \Tapp100\K are first-order heating and cooling transitions between the two phases. The inset shows a single crystal grown by vapor transport. The orange background corresponds to the LT \Ctm monoclinic phase.
			}
		\label{Fig:Props}
\end{figure}

The temperature dependent magnetic susceptibility of single crystalline \NbC upon cooling and warming is shown in Fig.~\ref{Fig:Props}\,(a) for fields parallel and perpendicular to the $c$ crystallographic axis. Whereas previous reports show that \NbC has hysteretic magnetization with a clear transition around \Tapp90\K\cite{kennedy_-nb3cl8_1992}, we find the susceptibility effectively vanishes for $T < 90$\K upon cooling. Hysteresis is observed, in that upon warming from the low-temperature (LT) phase to the HT phase, this transition occurs at a higher temperature than upon cooling. An analysis of the inverse susceptibility data for $T > 140$\K yields a Curie constant $C = 0.484\ \mathrm{emu \cdot K \cdot mol\ \mathit{f.u.}^{-1} \cdot Oe^{-1}}$ ($p_{eff}=1.97$, consistent with \Shalf) and a Weiss temperature of $\theta = -51.2$\,K. The fit is shown as a solid black line in Fig.~\ref{Fig:Props}\,(a), for $140\,\mathrm{K} \leq T \leq 300\,\mathrm{K}$ and is continued as a dashed line at lower temperatures. The values vary less than $5 \%$ between the two crystal directions, indicating any HT anisotropy is small. The upturns observed below \Tapp30\K account for $\approx 2 \%$ of the high-temperature spins, consistent with a small number of impurity spins or edge states. The fit shown below \Tapp90\K in Fig.~\ref{Fig:Props}\,(a) is to a Curie-Weiss law, $\chi = \frac{C}{T-\theta} + \chi_\mathrm{o}$ where $\chi_\mathrm{o} = 0.8 \times 10^{-4}\ \mathrm{emu \cdot Oe^{-1} \cdot {mol}}\ f.u.^{-1}$ and $\theta = -4$\K.

\begin{figure}[b!]
	\centering
		\includegraphics[width=8cm]{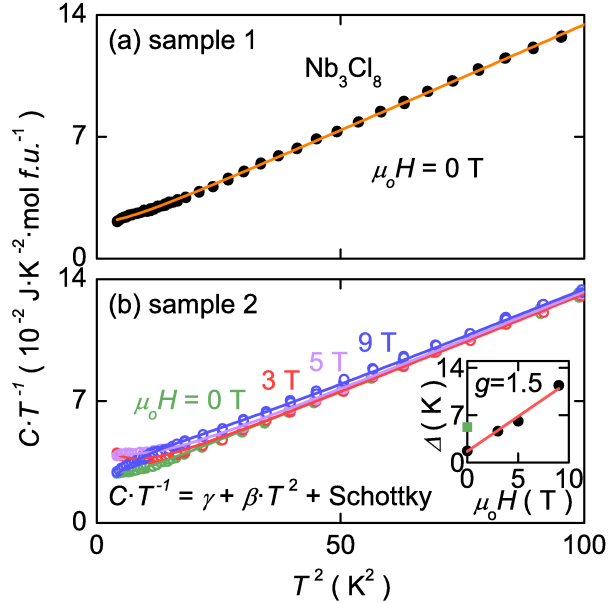}
			\caption{
			Low temperature $C_P \cdot T^{-1}$ versus $T^2$ data on two pieces (a,b) of the same batch of \NbC. Dots are data, lines are fits to the equation shown in (b). The values of $\gamma$ extracted from fits to the data are $\approx 40\%$ different for the two pieces of \NbC crystal, suggesting the origin of the $T$-linear component is from impurity spins. The inset in (b) shows the value of the Schottky gap as a function of field extracted from fits to data on sample 1 (green square) and sample 2 (black dots), which yield a $g$-factor of $1.5(1)$.
			}
		\label{Fig:HC}
\end{figure}

Fig.~\ref{Fig:Props}\,(b) shows the results of neutron diffraction rocking scans of the \NbC (001) reflection. Compared to the HT phase, the LT phase has a significant increase in the intensity of the (001) reflection, which tracks the transition between the LT and HT phases, with consistent hysteresis.

Heat capacity measurements on \NbC are shown in Fig.~\ref{Fig:Props}\,(c). The cooling (blue) and warming (red) transitions are consistent with the susceptibility and diffraction data. The amount of entropy associated with the transitions is obtained from $S(T) = \int^T_0 C_v \cdot T^{-1} dT$ assuming $C_v \approx C_p$, after a smooth curve to account for the background is subtracted. This results in an entropy change $\Delta S = 5\ \mathrm{J \cdot K^{-1} \cdot mol\ \mathit{f.u.}^{-1}}$, which is $85\,\%$ of the full entropy change of a two level system, $R \cdot ln(2) = 5.76\ \mathrm{J \cdot K^{-1} \cdot mol\ \mathit{f.u.}^{-1}}$. This may suggest the bulk has a singlet ground state with a gap to the first excited state.

An analysis of the low temperature $C_P \cdot T^{-1}$ versus $T^2$ data (Fig.~\ref{Fig:HC}), however, reveals a significant linear contribution to the specific heat, indicative of metallic behavior, despite \NbC being an insulator at all accessible temperatures. Measurements on two separate single crystal pieces yield two different linear contributions, $\gamma = 13\ \mathrm{and}\ 18\ \mathrm{mJ \cdot K^{-2} \cdot mol}\ f.u.^{-1}$ for samples 1 and 2 respectively [Fig.~\ref{Fig:HC}\,(a) and (b)]. The observation of varying values of the \textit{T}-linear contribution to the specific heat suggests its origin is not intrinsic but rather due to defects that also give rise to the Curie tail in low temperature susceptibility data [Fig.~\ref{Fig:Props}\,(a)]. To quantify this argument, the $C$($T$) data were fit to an equation accounting for the intrinsic sample behavior with a Schottky anomaly term for the low temperature impurity spins, $C \cdot T^{-1} = \beta \cdot T^2 + \gamma + (\frac{R}{T})(\frac{\Delta}{T})^2\frac{e^{\Delta / T}}{[1+e^{\Delta / T}]^2}$, where $\gamma$ is the $T$-linear contribution, $\beta$ is the lattice contribution, $R$ is the ideal gas constant, and $\Delta$ is the Schottky anomaly gap. A single value of $\beta$ and a single scale factor for the low temperature Schottky anomaly was used for all fields of \mbox{sample 2} [Fig.~\ref{Fig:HC}\,(b)]. The field dependence of the Schottky anomaly gap is shown in the inset to Fig.~\ref{Fig:HC}\,(b), which is used to extract a $g$-factor of $1.5(1)$ for the impurity spins contributing to the anomaly. We thus reason the $T$-linear specific heat of \NbC may originate from a distribution of broken singlets (giving rise to defect spins) that interact through the majority phase as in \BaCu\cite{zhou_spin_2011} and Sr$_2$CuO$_3$\cite{sologubenko_thermal_2000}.

\begin{figure}[t!]
	\centering
		\includegraphics[width=7.7cm]{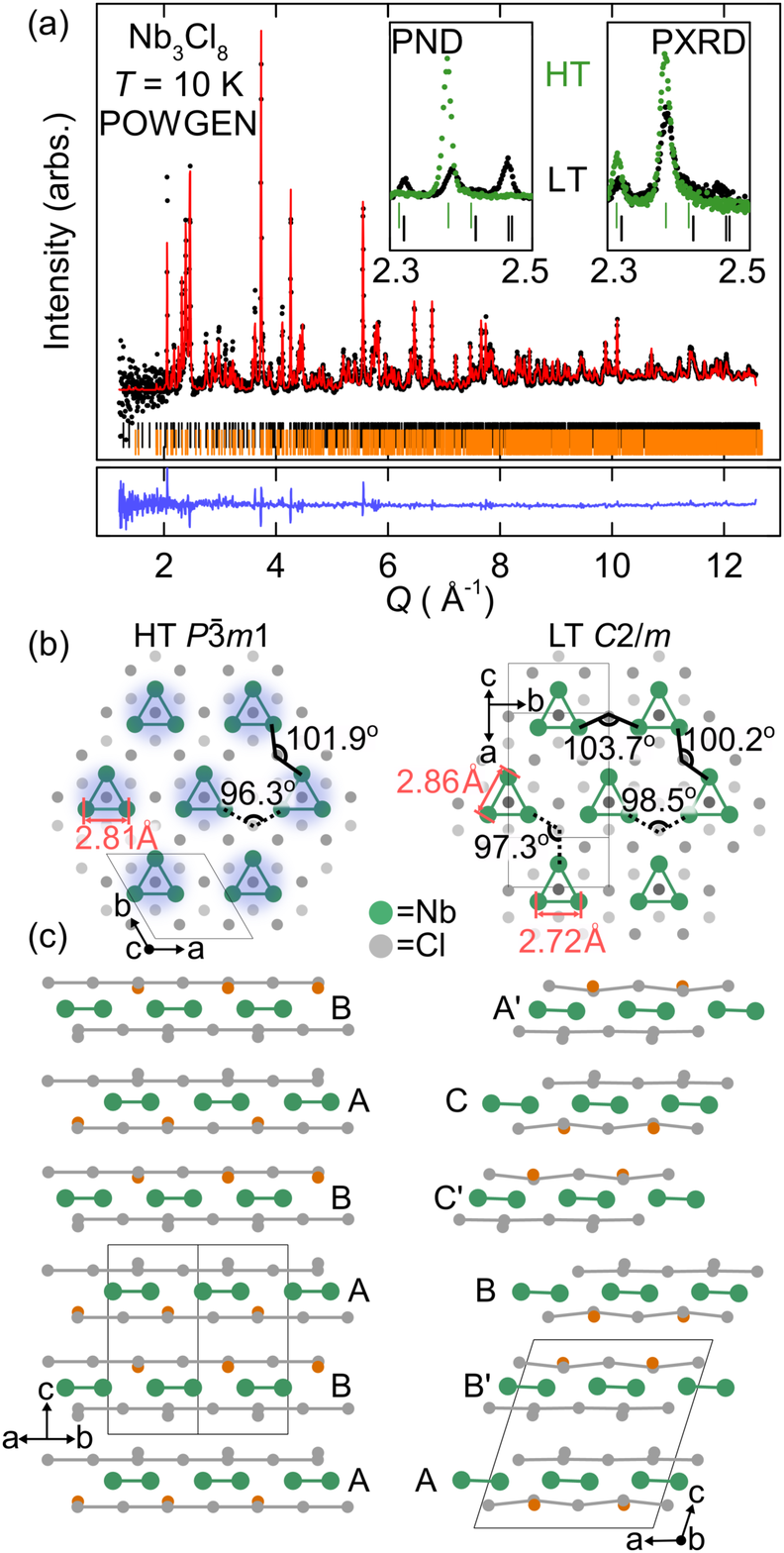}
			\caption{
			(a) Neutron diffraction at \Teq10\K indicates LT-\NbC adopts \Ctm space-group symmetry. Black dots are data, red line is calculated fit, blue line is difference, and tick marks are Bragg reflections for LT (black) and remnant HT (orange) phase \NbC. Insets show HT (green) and LT (black) data from neutron and \mbox{X-ray} diffraction at \Teq10\K and \Teq12\K, respectively, showing the split HT (202) trigonal reflection. HT and LT phase \NbC structures are shown perpendicular (b) and parallel (c) to \NbC layers. (b) Changes in intra-cluster Nb-Nb distances (red) and inter-cluster Nb-Cl-Nb bond angles (black) depict effect on intra-layer superexchange pathways. A shift in layer stacking (c) is observed from a HT -aaa- bilayer \mbox{(-AB-AB- layer)} to LT -abc- bilayer \mbox{(-AB$^\prime$-BC$^\prime$-CA$^\prime$- layer)} order. Orange Cl atoms highlight puckering of Cl atom layers which drive closest packing and stacking order rearrangement.
			}
		\label{Fig:LTm}
\end{figure}

\begin{table}[t!]
		\caption{Atomic parameters of LT-\NbC from Rietveld refinement to neutron data at \textit{T}~=~10\,K. The spacegroup is \Ctm with lattice parameters $a = 11.6576(1)$\,\AA{}, $b = 6.7261(1)$\,\AA{}, $c = 12.8452(1)$\,\AA{} and $\beta = 107.6087(2)^\textrm{o}$. All sites are fully occupied and thermal parameters were set equal for all atoms, with a refined value of $B_{iso} = 0.10(6)$. The fit quality is given by a $\textrm{R}_{wp} = 3.59$.}
	\begin{tabular*}{\columnwidth}{@{\extracolsep{\fill}}lclll}
		\hline
    \bf{Atom} & \textbf{Wyck. site} & \textbf{x} & \textbf{y} & \textbf{z} \\
		\hline
	Nb-1 & 4\textit{i} & 0.9734(10) & 1/2 & 0.2470(10) \\
	Nb-2 & 8\textit{j} & 0.1558(10) & 0.2980(10) & 0.2550(10) \\
	Cl-1 & 4\textit{i} & 0.7046(10) & 0 & 0.1231(10) \\
	Cl-2 & 4\textit{i} & 0.2157(10) & 0 & 0.1473(10) \\
	Cl-3 & 4\textit{i} & 0.1271(10) & 1/2 & 0.4013(10) \\
	Cl-4 & 4\textit{i} & 0.1239(10) & 0 & 0.3656(10) \\
	Cl-5 & 8\textit{j} & 0.9595(10) & -0.2490(10) & 0.1209(10) \\
	Cl-6 & 8\textit{j} & 0.3725(10) & 0.2606(10) & 0.3622(10) \\
		\hline
	\end{tabular*}
		\label{Tab:LT_Rietm}
\end{table}

Low temperature powder \mbox{X-ray} and neutron diffraction was used to determine the LT structure. The LT phase has a characteristic pattern of peak splitting, indicating a structural distortion to a \Ctm phase. Rietveld refinement to neutron diffraction data at \Teq10\K, shown in Fig.~\ref{Fig:LTm}\,(a), indicates a 16\% remnant, random HT phase stacking pattern in the LT structure. The resulting structure is summarized in Table~\ref{Tab:LT_Rietm} and change in symmetry unique octahedral Nb-Cl bond lengths in Table~\ref{Tab:Nb_Cl_dist}. Considerable stacking faults and the remnant HT phase stacking present in the LT phase warranted use of FAULTS\cite{casas_faults_2006} to model the LT structure with all thermal parameters restrained to be equal to minimize refinable parameters and yield the most statistically significant model. The primary effect of the phase transition on the \NbCC clusters is shown in Fig.~\ref{Fig:LTm}\,(b) and (c). The discrete \NbCC clusters ($2.81$\,{\AA} intra- and $3.93$\,{\AA} inter-cluster Nb-Nb bond distances) in the HT phase [Fig.~\ref{Fig:LTm}\,(b)] are characterized by ``molecular'' $C_{3v}$ point group symmetry and a \Shalf magnetic electron equally distributed over the entire cluster (blue shading). The clusters form a triangular lattice that are stacked in a \mbox{-AB-AB-} sequence. The transition to the LT phase [Fig.~\ref{Fig:LTm}\,(b) and (c)] results in removal of the threefold rotational symmetry of the HT clusters, so that only a \Ctm $a$-$c$ mirror plane remains. The result is a scissoring of the clusters, whereby the HT phase equilateral Nb$_3$ triangular cluster becomes isosceles, resulting in one decreased (2.72\,\AA{}) and two increased (2.86\,\AA{}) Nb-Nb bond lengths and two inequivalent NbCl$_6$ octahedra, consistent with the SOJT effect. The transition modulates inter-cluster Nb-Cl-Nb superexchange pathways by decreasing and increasing these bond angles [Fig.~\ref{Fig:LTm}\,(b)] resulting in a pseudo-one-dimensional state. Further, accompanying the phase transition are changes to the nearest-neighbor (NN) inter- and intra-cluster distances, here defined as the distance between the geometric centroid of Nb$_3$ cluster triangles. While the intra-layer cluster distances are shorter and change from a single NN distance of 6.7457\,{\AA} (HT) to 6.7294\,{\AA} and 6.7261\,{\AA} (LT)---suggestive of 1D chains---the change in NN inter-layer distance is greater, from 7.3585\,{\AA} (HT) to 7.1887\,{\AA} (LT), consistent with inter-layer singlet formation. The cluster scissoring has a pronounced effect on the interlayer stacking arrangement. In the HT phase, Cl atoms at bilayer edges are flat triangular layers, resulting in a simple closest packing of chlorines from the top of one bilayer with the bottom of the next and thus, a \mbox{-AB-AB-} arrangement. In the LT phase, the scissoring motion results in a buckling of the chlorine layers as highlighted in Fig.~\ref{Fig:LTm}\,(c), where crystallographically equivalent orange Cl atoms emphasize the Cl-layer crimping distortion---the lowest interfacial energy between these now staggered layers requires a shift in stacking to a \mbox{-AB$^\prime$-BC$^\prime$-CA$^\prime$-} pattern. Here, A/A$^\prime$, B/B$^\prime$, \textit{etc} is used to relate the LT structure to the former HT (AB-designated) layers. The resultant shift in inter- and intra-layer structure, however, is not concomitant with long-range magnetic order. The observation of a structural distortion by powder \mbox{X-ray} diffraction suggests the increased intensity of (001) neutron Bragg diffraction [Fig.~\ref{Fig:Props}\,(b)] has a structural---not magnetic---origin. Changes in the measured single crystal neutron diffraction data are consistent solely with nuclear diffraction associated with the \Ptm to \Ctm phase transition.

\begin{figure*}[t!]
	\centering
		\includegraphics[width=13cm]{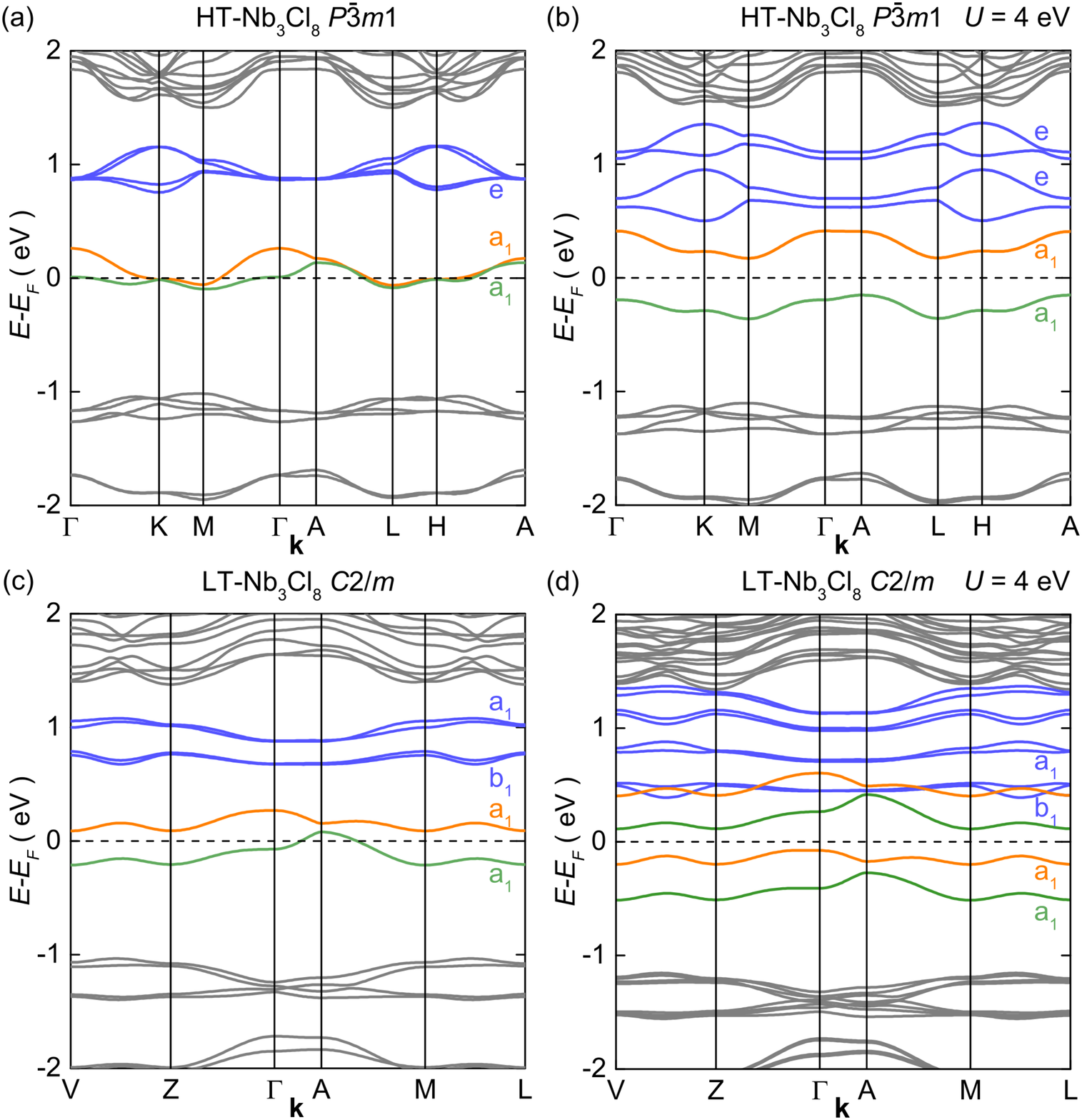}
			\caption{
			Band structure calculations on the HT [(a) and (b)] and LT [(c) and (d)] structures of \NbC. Calculations in (a) and (c) are performed without spin polarization (SP), spin-orbit coupling (SOC), or a Hubbard $U$. Calculations in (b) and (d) are performed with SP, SOC, and Hubbard~$U = 4$\,eV. The transition observed in \NbC is driven by a second-order Jahn-Teller distortion driven by interaction of (a) HT phase $e$ (blue) and $a_1$ orbitals. The HT phase possesses a bonding (green) and anti-bonding (orange) pair of states from the non-degenerate ($a_1$) valence bands at the $\Gamma$- point from an interaction between interlayer clusters. The HT phase $e$ orbitals are split upon transition into the (c) LT state. To reproduce the experimentally observed insulating behavior in the HT calculation, the Hubbard $U$ and SOC must be included (b), which splits both the non-degenerate $a_1$ and $e$ bands. For comparison, inclusion of a Hubbard $U$ in the LT calculations is shown in (d) (see ESI$^\dag$). The special points of the Brillouin zone in both the HT and LT calculations are listed in the ESI$^\dag$.
			}
		\label{Fig:BANDSm}
\end{figure*}

{
\renewcommand\arraystretch{1.1}
\begin{table}[t!]
		\caption{Changes in the niobium-chlorine bond lengths of symmetry unique NbCl$_6$ octahedra in \NbCC clusters between the HT and LT \NbC phases. Bond lengths are separated and labeled by unique Cl atoms within an octahedra. Duplicate lengths within a NbCl$_6$ octahedra are labeled as such. Nb$_A$ is the apical (\textit{i.e.} non-scissored) Nb ion, Nb$_B$ represent the two ``scissored'' Nb ions. All distances are in Angstroms (\AA{}).}
	\begin{tabular*}{\columnwidth}{@{\extracolsep{\fill}}lrlr}
		\hline 
		    \multicolumn{2}{c}{\textbf{HT} (\Ptm)} & \multicolumn{2}{c}{\textbf{LT} (\Ctm)} \\
		\hline
			\multirow{2}{*}{Nb-Cl$_A$} & \multirow{2}{*}{2.437(2)} & Nb$_A$-Cl$_{A1}$ & 2.483(13) \\
			 &  & Nb$_B$-Cl$_{A2}$ & 2.423(13) \\
		\hlinewd{0.2pt}
			\multirow{3}{*}{Nb-Cl$_B$} & \multirow{3}{*}{2 $\times$ 2.530(1)} & Nb$_A$-Cl$_{B1}$ & 2 $\times$ 2.552(11) \\
			 &  & Nb$_B$-Cl$_{B2}$ & 2.548(10) \\
			 &  & Nb$_B$-Cl$_{B3}$ & 2.497(13) \\
		\hlinewd{0.2pt}
			\multirow{3}{*}{Nb-Cl$_C$} & \multirow{3}{*}{2 $\times$ 2.464(1)} & Nb$_A$-Cl$_{C1}$ & 2 $\times$ 2.407(12) \\
			 &  & Nb$_B$-Cl$_{C2}$ & 2.435(12) \\
			 &  & Nb$_B$-Cl$_{C3}$ & 2.371(12) \\
		\hlinewd{0.2pt}
			\multirow{2}{*}{Nb-Cl$_D$} & \multirow{2}{*}{2.640(1)} & Nb$_A$-Cl$_{D1}$ & 2.513(13) \\
			 &  & Nb$_B$-Cl$_{D2}$ & 2.646(10) \\
		\hline 
	\end{tabular*}
		\label{Tab:Nb_Cl_dist}
\end{table}
}

Density functional theory (DFT) band structure calculations for \NbC, shown in Fig.~\ref{Fig:BANDSm}, were used to gain insight into the nature of the magneto-structural transition. In agreement with previous calculations for discrete \NbCC clusters\cite{miller_solid_1995}, the HT \NbC band structure confirms the highest occupied molecular orbital (HOMO) of each cluster is a non-degenerate $a_1$ orbital that is far in energy from the excited state orbitals and therefore, not conventionally (\textit{i.e.} to first-order) Jahn-Teller active. The bands corresponding to each \NbCC cluster HOMO [green and orange band, one from each of two clusters per unit cell, Fig.~\ref{Fig:BANDSm}\,(a)] are split at the $\Gamma$-point from orbital overlap between the magnetic electrons in adjacent clusters resulting in the formation of a bonding/anti-bonding pair of states. The band structure calculation of the LT phase, Fig.~\ref{Fig:BANDSm}\,(c), shows shifted bands with respect to the HT phase. Notably, the HT bonding/antibonding states [from Fig.~\ref{Fig:BANDSm}\,(a)] have an increased gap and the degenerate $e$ states have been raised and lowered in energy (\textit{i.e.} orbital degeneracy was broken) upon entering the LT phase. Breaking the degeneracy of the first excited states upon entering in the LT phase is consistent with a second-order Jahn-Teller distortion from orbital mixing. The calculations in Fig.~\ref{Fig:BANDSm}\,(a) and (c) have no spin-polarization (SP), spin-orbit coupling (SOC), or electronic interaction \textit{via} a Hubbard $U$. Under these conditions, HT \NbC is clearly predicted to have a finite density of states at the Fermi level and thus be metallic, inconsistent with resistivity measurements. This demonstrates the importance of correlations in producing the observed behavior of this compound. Intriguingly, we find that solely SOC or solely a Hubbard $U$ (onsite electron-electron interaction), are not sufficient to produce an insulator in the HT phase (see ESI$^\dag$). Including both SOC and $U$ simultaneously, Fig.~\ref{Fig:BANDSm}\,(b), is sufficient to produce an insulator. In this calculation, Nb moments were assumed to align with the crystallographic $c$ axis, and be oriented antiferromagnetically between layers with a total moment of $\approx 0.5\,\mu_{\mathrm{B}}$ per cluster in the HT phase. Given the limitations of DFT in describing correlated materials, this semi-quantitative agreement is reasonable. In contrast, LT \NbC is almost predicted to be insulating by DFT, even in the absence of SOC and a Hubbard $U$, Fig.~\ref{Fig:BANDSm}\,(c). This demonstrates the importance of changes in interlayer interactions, as the splitting of the HT $a_1$ bands induces the SOJT effect, driving this material toward an insulating state [of course inclusion of SOC and $U$ increase the insulating gap, Fig.~\ref{Fig:BANDSm}\,(d)].

\section{Discussion} 

The data demonstrate a first-order phase transition from a triangular lattice paramagnet with antiferromagnetic correlations to a non-magnetic state. The transition temperature \Tapp90\K is higher than the Weiss temperature ($\theta$), a measure of the magnetic interaction strength. Typically a material, absent of any magnetic frustration effects, will form a magnetically long-range ordered state with $T_{order} = |\theta|$, with the kind of ordering (ferro- or antiferromagnetic) dependent on the sign of the Weiss temperature. In \NbC, the onset of the SOJT effect well above the temperature at which in-plane magnetic correlations become strong suggests an inherent instability of the system. The effects of magnetic frustration observed in another triangular lattice cluster-magnet based material, \LiZn\cite{sheckelton_possible_2012,sheckelton_local_2014,sheckelton_electronic_2014,mourigal_molecular_2014}, is proposed to result from itinerant, plaquette charge order\cite{chen_spin_2014} that does not form a long-range ordered state. In \LiZn, as opposed to \NbC, stability of the frustrated magnetic state can be due to the non-magnetic Li/Zn ions suppressing inter-layer magnetic interactions predicted by band structure calculations (Fig.~\ref{Fig:BANDSm}). In addition, previous reports suggest the Nb-Cl bonding within \NbCC clusters has strong ionic character\cite{kennedy_experimental_1996} as opposed to highly covalent bonding, which might give insight into the lack of stability of the frustrated magnetic state---the ionic character of the cluster bonds is more susceptible to a SOJT effect and subsequent structural distortion due to the low energy barrier to overcome.

\begin{figure}[t!]
	\centering
		\includegraphics[width=13cm]{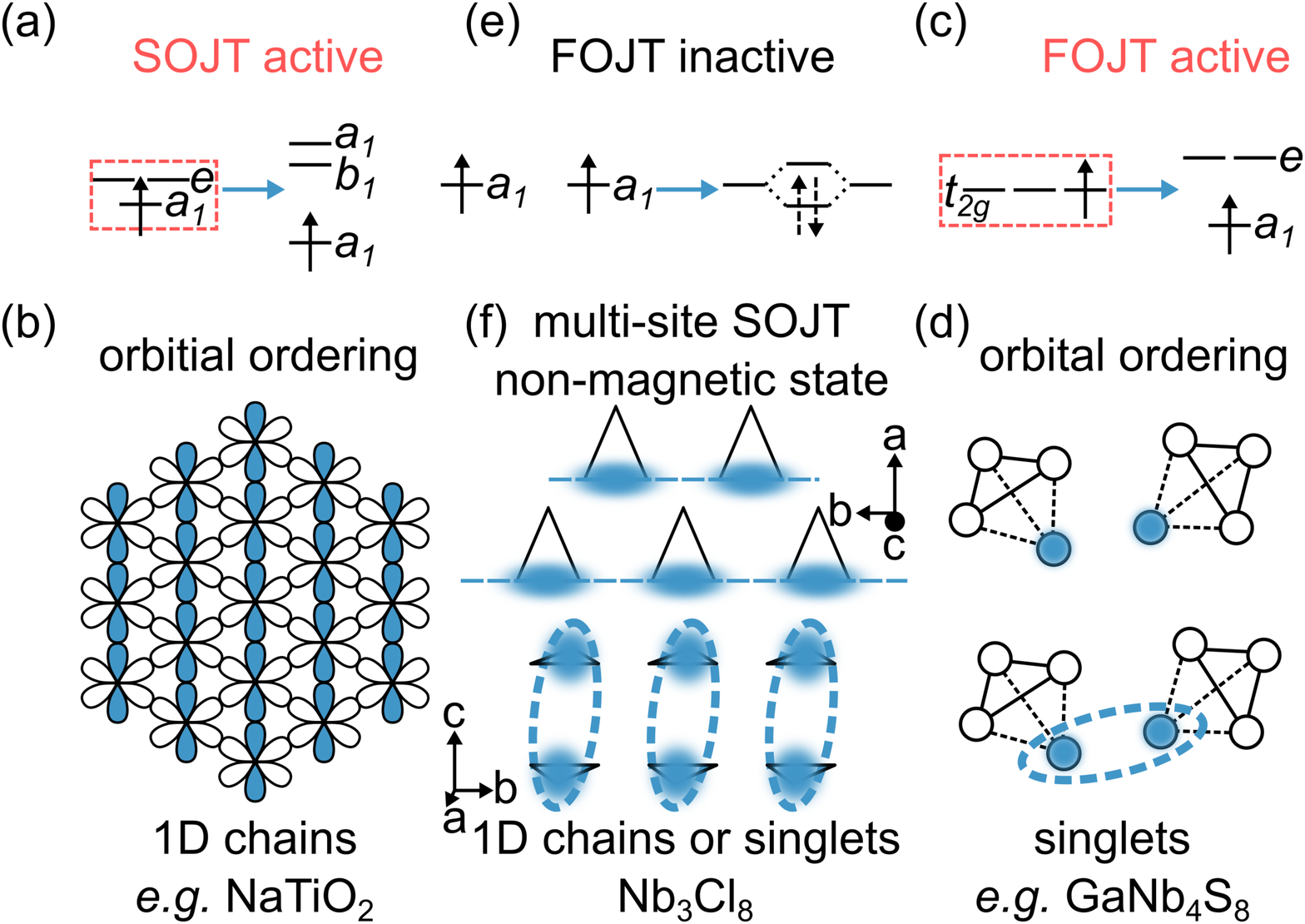}
			\caption{
			Subtle electronic and magnetic interactions can cause materials to avoid frustration. Simplified band pictures show Jahn-Teller (JT) active distortions upon entering low temperature phases in (a) NaTiO$_2$ and (c) \GaNbS. (e) The HT phase of \NbC has a multi-site interaction between $a_1$ orbitals on adjacent, inter-layer \NbCC clusters. The distortion to the LT phase is driven by a SOJT effect, despite the $\approx~1$\,eV energy gap, and thus, vanishingly small thermal population of the degenerate $e$ orbitals. Magnetic order in various materials is achieved (b) by the formation of 1D chains (in plane Ti-$d$ orbitals shown) \textit{via} orbital ordering (OO) in NaTiO$_2$, (d) OO and singlet formation in \GaNbS. Further neutron scattering experiments are required to determine either (f) 1D chains or singlets as the magnetic ground state in \NbC.
			}
		\label{Fig:Cartoon}
\end{figure}

We compare the peculiar structural distortion observed in \NbC with other anomalous magnetism in materials and 2D VdW structures. It has been suggested that the observed properties of the triangular lattice antiferromagnet NaTiO$_{2}$\cite{clarke_synthesis_1998} can be explained by a one-dimensional-like behavior\cite{pen_orbital_1997,dhariwal_orbital_2012} arising from a symmetry lowering structural distortion, where energetically similar orbitals are thermally populated at high temperatures, Fig.~\ref{Fig:Cartoon}\,(a). Upon cooling, a distortion relieves the frustration by orbital ordering of the Ti$^{3+}$ cations to form 1D chains, Fig.~\ref{Fig:Cartoon}\,(b). Analogous behavior is observed in other compounds with single ions as the basic magnetic unit, including NaVO$_2$\cite{mcqueen_successive_2008} and LiVO$_2$\cite{tian_single_2004}. A similar phenomenology is found in cluster-based Lacunar spinels such as \GaNbS\cite{pocha_crystal_2005}. There, a first-order Jahn-Teller distortion due to a single electron in a $t_{\mathit{2}g}$ triply degenerate orbital set, Fig.~\ref{Fig:Cartoon}\,(c), results in the localization of the unpaired electron on a single Nb atom in the Nb$_4$S$_4$ cores, Fig.~\ref{Fig:Cartoon}\,(d). This ordering is concomitant with a drop in magnetism, thought to result from the formation of spin-singlets between two clusters\cite{waki_spin_2010} \textit{via} interaction of adjacent-cluster Nb atoms [Fig~\ref{Fig:Cartoon}\,(d)]. Singlet formation born from orbital ordering is also seen in other potentially frustrated systems such as \BaCu\cite{nakatsuji_spin-orbital_2012} and \BaCo\cite{shirata_experimental_2012}. In all of these cases, the orbital degree of freedom plays a crucial role in driving the structural distortion that relieves magnetic frustration and results in either magnetic order or a singlet state.

\NbC exhibits similar physics despite the absence of an orbital degeneracy (\textit{i.e.} first-order Jahn-Teller instability) or significant thermal population of the first excited state. The results of band structure calculations on LT-\NbC [Fig.~\ref{Fig:BANDSm}\,(c)] indicate second-order Jahn-Teller (SOJT) like orbital mixing, since upon entering the LT state, HT degenerate $e$ bands [blue bands, Fig.~\ref{Fig:BANDSm}\,(a)] break their degeneracy and a scissoring of the Nb$_3$ triangles is expected from the SOJT effect. This is surprising given the $\approx~1$\,eV energy gap to the degenerate $e$ states in the HT band structure, which are expected to have little-to-no thermal population at \textit{T}~$\approx$~90\K. The HT band structure calculation predicts an electronic interaction by orbital overlap between magnetic electrons in adjacent \NbC layers---yielding effective bonding/antibonding states---possibly inducing the SOJT effect and is suggestive of an inter-layer singlet ground state in the LT phase. Such a singlet ground state would be highly unusual given the large distance between Nb ions in adjacent layers, although magnetic exchange interactions are possible on such length scales, as in \mbox{$\alpha$-RuCl$_3$}\cite{sandilands_scattering_2015, sears_magnetic_2015, banerjee_proximate_2016}, which harbors an out of plane magnetic interaction and similar stacking faults as LT \NbC. Alternatively, the concurrent change in intra-layer, inter-cluster Nb-Cl-Nb bond angles upon entering the LT phase could result in a intra-layer \Shalf 1D chain singlet state. While at present, the LT magnetic ground state cannot be unambiguously determined, singlets that lie in between adjacent layers or a intra-layer \Shalf 1D chain appear to be the most likely candidates. Indeed, while no in-plane dimerization is observed, dimerization \textit{via} electronic interaction in the $c$-direction exists even in the HT phase.

The DFT calculations affirm the HOMO in each HT \NbCC cluster are non-degenerate ($a_1$), so the transition in \NbC cannot be ascribed solely to single-site orbital ordering. Related phenomena are observed in two-dimensional organic salts where charge ordering is the cause of the metal-insulator transition as in $\theta$-(BEDT-TTF)$_2$RbZn(SCN)$_4$\cite{abdel-jawad_anomalous_2010}. The observation of a dielectric response in the geometrically frustrated dimer Mott insulator $\kappa$-(BEDT-TTF)$_2$Cu$_2$(CN)$_2$\cite{miyagawa_charge_2000} suggests charge order plays a role in the putative spin liquid state. Devoid of a screw-axis, the HT \Ptm space group is symmorphic, and therefore the formation of a gapped, trivial magnetic ground state is allowed without a structural distortion in \NbC\cite{oshikawa_commensurability_2000,parameswaran_topological_2013}. Instead, the observation of the SOJT effect and subsequent non-magnetic ground state in the presence of non-degenerate HOMO bands and unpopulated degenerate $e$ bands reveals that the electronic interaction [Fig.~\ref{Fig:Cartoon}\,(e)] due to orbital overlap plays a crucial role in the avoidance of a magnetically frustrated, classically degenerate ground state. Estimates of the change in entropy and enthalpy from the specific heat measurements in Fig.~\ref{Fig:Props}\,(c) suggest the first-order transition is entropically driven, by the enhanced spin entropy of the frustrated magnetic HT phase over that of a supposed spin singlet LT phase. The apparent lack of long-range magnetic order in LT-\NbC is a natural consequence of the two likely magnetic ground state scenarios---formation of a spin-ladder network of inter-layer singlets or intra-layer 1D chains, illustrated in Fig.~\ref{Fig:Cartoon}\,(f). The first-order nature of the transition prevents estimation of the low temperature magnetic exchange interactions (\textit{e.g.} the magnitude of the spin gap) based on the HT phase properties. The deeply suppressed magnetic response in the LT phase implies a gap in the LT state, $\Delta \approx 150-650$\,K (see ESI$^\dag$), that is significantly larger than the HT Weiss temperature, but not inconsistent with the $\approx 200 $\,meV splitting of the $a_1$ orbitals at the $\Gamma$-point in the DFT calculations. Further spectroscopic investigations, such as inelastic neutron scattering, are required to elucidate the exact nature of the LT magnetic ground state.

Lastly, the discovery of extraordinary physical properties of graphene\cite{novoselov_electric_2004,novoselov_roadmap_2012} and other 2D VdW materials---such as transition metal dichalcogenides\cite{wang_electronics_2012,chhowalla_chemistry_2013,qian_quantum_2014,ali_large_2014}---has prompted materials scientists to feverishly search for new systems with properties suitable to, \textit{e.g.}, fabricate technologically useful devices. Like \NbC, these materials are (strongly) covalently bonded in two dimensions while adhesion in the third dimension is controlled by comparatively weak VdW forces. This weak bonding in one dimension allows for unparalleled control over heterostructure fabrication possibilities due to the facile nature by which differing interlayers can be combined\cite{geim_van_2013,liu_van_2016} to create nano-scale electronic devices such as precisely tunable transistors\cite{radisavljevic_single-layer_2011,lee_atomically_2014} or LEDs\cite{withers_light-emitting_2015}. \NbC is a rare example of a material, despite its absence of ``molecular'' (\NbCC) HOMO degeneracy and energetically isolated HOMO bands, that undergoes a rearrangement of layer stacking at low temperature due to inter-layer magnetic interactions. Our insight gained into the SOJT-like effect in \NbC, which drives the puckering of interfacial VdW Cl layers and forces the shift in stacking arrangement, opens up possibilities for low-temperature device applications where controllable structural changes are desired. The ability for interactions in the VdW-layer-direction to control crystal structure can be combined with other 2D VdW layers to create novel devices.

\section{Conclusions}

In short, \NbC highlights the interplay between charge, orbital, and spin degrees of freedom and the approach to a non-entropic state as $T \rightarrow 0$. While the space group symmetry does not preclude a singlet ground state in the symmorphic high temperature structure with two valence electrons\cite{rondinelli_structure_2011,watanabe_lieb-schultz-mattis_2015}, such a state is apparently not energetically favorable. At low temperature, on the other hand, the strongly reduced susceptibility and inter-layer HOMO overlap suggested by DFT point to a singlet ground state of the valence bond crystal variety. While the structural transition must be energetically favorable for low-$T$ magnetism, the second-order Jahn-Teller driven structural rearrangement must be costly. The transition is thus analogous to a spin-Peierls transition where an energetically unfavorable structural dimerization is induced to lift spin degeneracy. For \NbC, a preference to avoid frustrated triangular lattice antiferromagnetism may even drive covalent bonding. Unheard of in conventional local moment magnetism, this scenario may not be uncommon for 4$d$ or 5$d$ magnetic materials, and has recently been proposed as an explanation for the behavior of Li$_2$RuO$_3$ and related iridium-based materials\cite{kimber_valence_2014}. More generally, that such a change in stacking of a Van-der-Waals material can occur near liquid nitrogen temperature is remarkable, and implies that designer 2-D heterostructures may be engineered to undergo similar phase transitions for a variety of potential applications. Akin to the wide breadth of physical phenomena discovered in graphene, the Van-der-Waals layered structure of \NbC may also yield novel physics as the first mono-layer geometrically frustrated magnet.

\section{Acknowledgements}
This research was supported by the US Department of Energy, Office of Basic Energy Sciences, Division of Materials Sciences and Engineering under Award DE-FG02-08ER46544 to The Institute for Quantum Matter at JHU. Use of the Advanced Photon Source at Argonne National Laboratory was supported by the U. S. Department of Energy, Office of Science, Office of Basic Energy Sciences, under Contract No. DE-AC02-06CH11357. Utilization of the mail-in program at POWGEN, Spallation Neutron Source, ORNL was sponsored by the Scientific User Facilities Division, Office of Basic Energy Sciences, U.S. Department of Energy. TMM acknowledges support from the David and Lucile Packard Foundation and the Sloan Research Fellowship. The authors would like to thank O. Tchernyshyov and A. Turner for useful discussions, L. Harriger for his support on the SPINS spectrometer, and S. Lapidus for his support on beamline 11-BM.

\newpage
\begin{center}
\noindent\LARGE{\textbf{Supplementary Information:\\ Rearrangement of Van-der-Waals Stacking and Formation of a Singlet State at \Teq90\K in a Cluster Magnet}}\\
\vspace{0.5in}
\noindent\large{John P. Sheckelton,\textit{$^{a,b}$} Kemp W. Plumb,\textit{$^{b}$} Benjamin A. Trump,\textit{$^{a,b}$} Collin L. Broholm,\textit{$^{b,c,d}$} \\ and Tyrel M. McQueen\textit{$^{a,b,c,*}$}}
\end{center}

\FloatBarrier
\setcounter{figure}{0}
\renewcommand{\thefigure}{S\arabic{figure}}
\def\lofname{}

\setcounter{table}{0}
\renewcommand{\thetable}{S\arabic{table}}
\def\lofname{}

\setcounter{section}{0}
\renewcommand{\thesection}{S\arabic{section}}
\def\lofname{}

\section{Synchrotron powder diffraction analysis}

\begin{figure}[h!]
	\centering
		\includegraphics[width=4in]{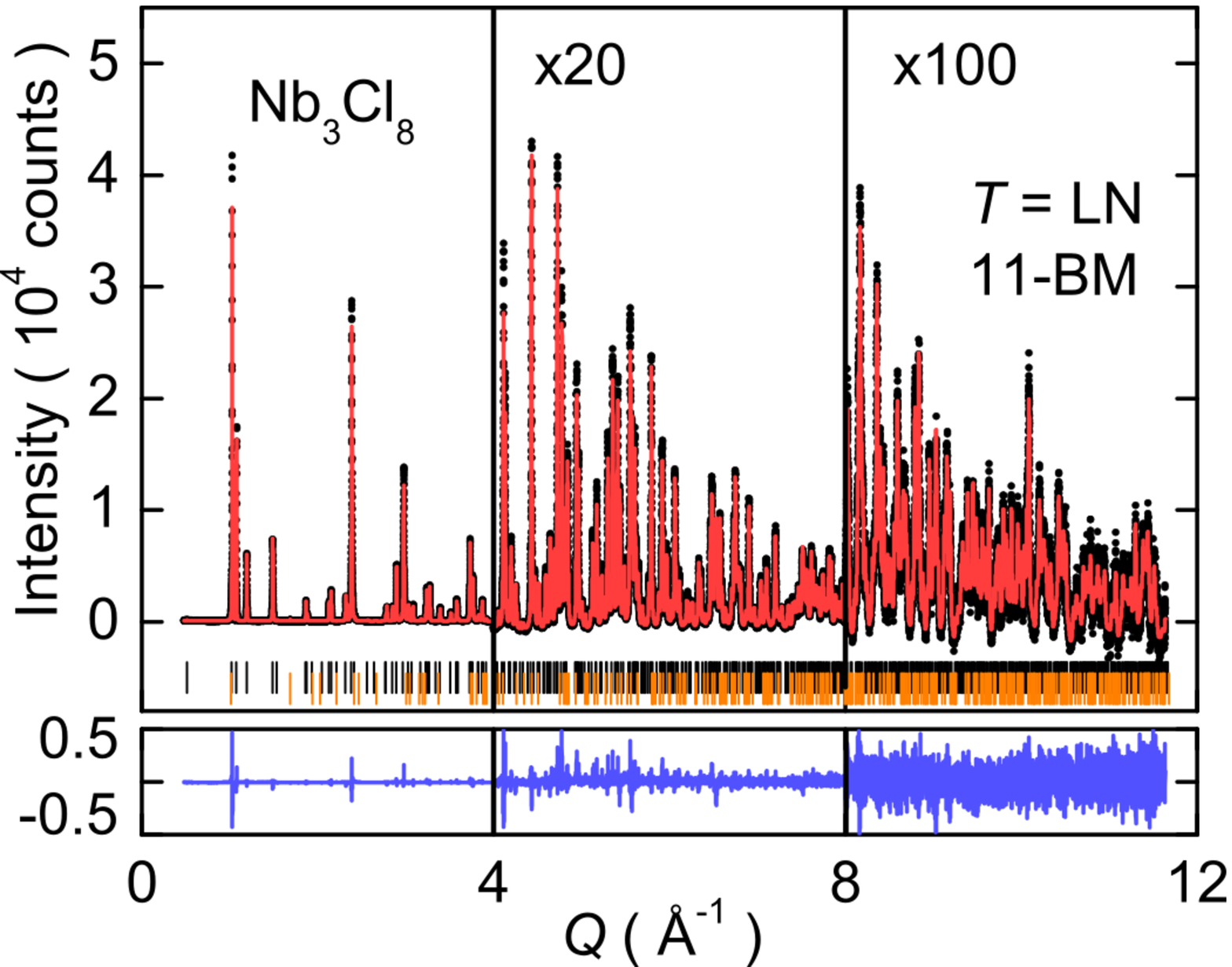}
			\caption{\NbC Rietveld analysis of synchrotron \mbox{X-ray} diffraction data at liquid nitrogen (LN) temperatures. Black dots are data, red line is the calculated fit, blue line is the difference, tick marks are Bragg reflections for \NbC (black) and a minor impurity, NbOCl$_2$ (orange, 0.5(2) wt\%). The higher $Q$ data are multiplied $\times\!20$ ($4 \geq Q \geq 8$) and $\times\!100$ ($8 \geq Q \geq 12$) to highlight the quality of the fit.}
		\label{Fig:LT}
\end{figure}

Table~\ref{Tab:HT_Riet} is a summary of Rietveld refinement parameters to \Teq300\K synchrotron \mbox{X-ray} powder diffraction data shown in main text Fig.~1(a). Low temperature synchrotron \mbox{X-ray} powder diffraction data near liquid nitrogen temperatures shows no signs of symmetry lowering distortions despite a temperature reading of approximately 80\,K near the sample. Rietveld refinement of the HT (\Ptm) structure to the data near liquid nitrogen temperatures is shown in Fig.~\ref{Fig:LT} and is summarized in Table~\ref{Tab:LT_Riet}. 80\,K is far below the first-order phase transition in \NbC ($\approx 90$\,K)---the absence of any \NbC LT phase, therefore, is attributed to sample heating from the 30\,KeV ($\lambda \approx 0.41$\,\AA{}) incident \mbox{X-ray} beam. As is the case with the \Teq300\K data, fitting the low temperature data to lower symmetry models (\textit{P}$\bar{3}$, \textit{P}3\textit{m}, \textit{P}3, and \Ctm) results in no improvement of the fit, either quantitatively or qualitatively.

\begin{table}[t!]
		\caption{
		\NbC Rietveld refinement to synchrotron X-ray ($\lambda = 0.4132$\,\AA) powder diffraction data at \textit{T}~=~300\,K, shown in main text Fig.~\,1(a). The spacegroup is \Ptm(164) with lattice parameters $a = 6.74566(3)$\,\AA{}, $c = 12.28056(7)$\,\AA{}, $ \alpha = \beta =90^\textrm{o}$, and $\gamma = 120^\textrm{o}$. All sites are fully occupied, cell volume was calculated to be $V = 483.947(6)$\,\AA$^3$, and atomic displacement parameters were freely refined for all atoms. The fit quality is given by $\textrm{R}_{wp} = 11.34\,\%$, $\textrm{R}_{p} = 9.03\,\%$, and $\chi^2 = 3.286$. Errors are computed statistical uncertainties.
		}
	\begin{tabular*}{0.9\columnwidth}{@{\extracolsep{\fill}}lcllll}
		\hline
    \bf{Atom} & \textbf{Wyckoff position} & \textbf{x} & \textbf{y} & \textbf{z} & \textbf{U$_{iso}$(\AA$^2$)} \\
		\hline
	Nb & 6\textit{i} & 0.52763(2) & 0.05525(5) & 0.24577(5) & 0.00473(6) \\
	Cl-1 & 2\textit{d} & 2/3 & 1/3 & 0.0978(2) & 0.0045(4) \\
	Cl-2 & 2\textit{d} & 1/3 & 2/3 & 0.3555(2) & 0.0085(5) \\
	Cl-3 & 6\textit{i} & 0.6696(2) & -0.1652(1) & 0.1350(1) & 0.0069(3) \\
	Cl-4 & 6\textit{i} & 0.3374(2) & 0.1687(1) & 0.3835(1) & 0.0103(3) \\
		\hline
	\end{tabular*}
		\label{Tab:HT_Riet}
\end{table}

\begin{table}[t!]
		\caption{
		\NbC Rietveld refinement to synchrotron X-ray ($\lambda = 0.4132$\,\AA) powder diffraction data at \textit{T}~$\approx$~90\,K, shown in Fig.~\ref{Fig:LT}. The spacegroup is \Ptm(164) with lattice parameters $a = 6.73255(2)$\,\AA{}, $c = 12.22222(5)$\,\AA{}, $ \alpha = \beta =90^\textrm{o}$, and $\gamma = 120^\textrm{o}$. All sites are fully occupied, cell volume was calculated to be $V = 479.777(4)$\,\AA$^3$, and atomic displacement parameters were freely refined for all atoms. The fit quality is given by $\textrm{R}_{wp} = 9.31\,\%$, $\textrm{R}_{p} = 7.61\,\%$, and $\chi^2 = 2.141$. Errors are computed statistical uncertainties.
		}
	\begin{tabular*}{0.9\columnwidth}{@{\extracolsep{\fill}}lcllll}
		\hline
    \bf{Atom} & \textbf{Wyckoff position} & \textbf{x} & \textbf{y} & \textbf{z} & \textbf{U$_{iso}$(\AA$^2$)} \\
		\hline
	Nb & 6\textit{i} & 0.52746(2) & 0.05493(4) & 0.24622(4) & 0.00182(4) \\
	Cl-1 & 2\textit{d} & 2/3 & 1/3 & 0.0981(2) & 0.0024(3) \\
	Cl-2 & 2\textit{d} & 1/3 & 2/3 & 0.3560(1) & 0.0041(3) \\
	Cl-3 & 6\textit{i} & 0.6709(1) & -0.16456(6) & 0.13528(7) & 0.0023(2) \\
	Cl-4 & 6\textit{i} & 0.3359(1) & 0.16793(6) & 0.38192(8) & 0.0051(2) \\
		\hline
	\end{tabular*}
		\label{Tab:LT_Riet}
\end{table}

\FloatBarrier

\section{DFT band structure calculations}

\begin{figure}[h!]
	\centering
		\includegraphics[width=\columnwidth]{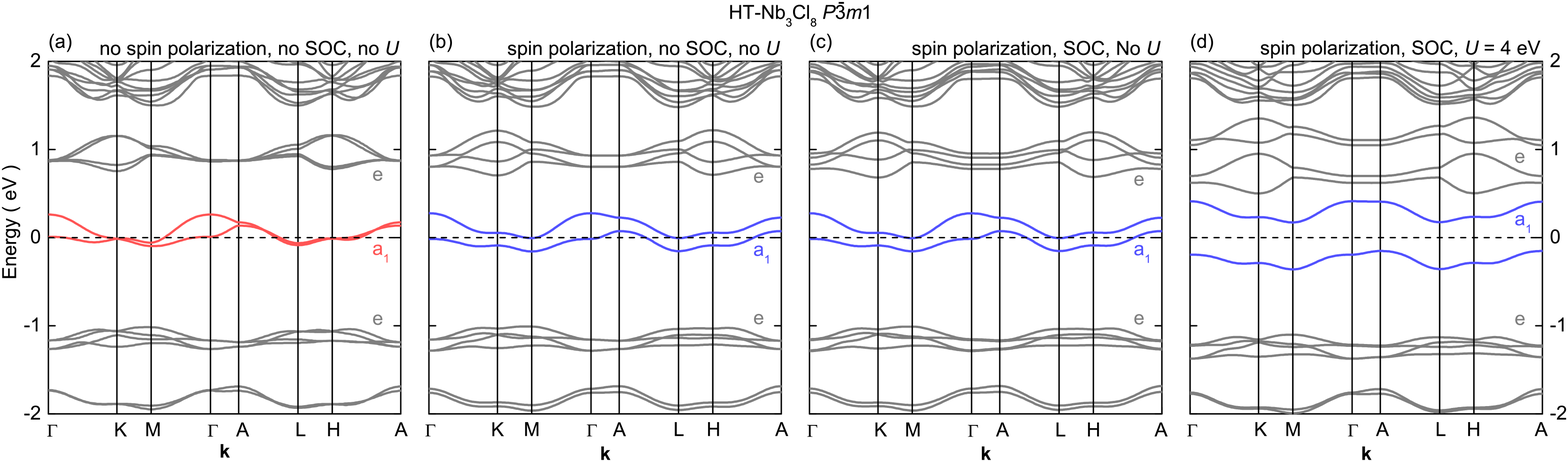}
			\caption[Density functional theory band structure calculations on the high temperature structure of \NbC.]{Band structure calculations on HT-\NbC with (a) no spin polarization (SP), spin-orbit coupling (SOC), or Hubbard $U$, (b) SP but no SOC/$U$ (c) SP and SOC but no $U$, and (d) SP, SOC, and $U = 4$\,eV. The red highlighted bands in (a) are the highest occupied $a_1$ orbitals, which are split at the $\Gamma$-point from the formation of a bonding/anti-bonding pair and are far in energy ($\approx1$ eV) from the degenerate $e$ orbitals. In (b)-(d), the total splitting of the $a_1$ band (blue) is due to antiferromagnetic order. The spin polarized calculations are initiated with the application of an applied field to break spin degeneracy. This results in calculations (b)-(d) converging with antiferromagnetic order where \NbCC cluster spins are oriented in the crystallographic $c$-axis. In (d), the antiferromagnetic band splitting is exacerbated by the electronic interaction induced by the Hubbard $U$.}
		\label{Fig:BANDS}
\end{figure}

Calculations on the high temperature (HT) structure of \NbC  were performed without spin-polarization [Fig.~\ref{Fig:BANDS}\,(a)], with spin-polarization [Fig.~\ref{Fig:BANDS}\,(b)], with spin-orbit coupling [Fig.~\ref{Fig:BANDS}\,(c)], and with a Hubbard $U$ term [Fig.~\ref{Fig:BANDS}\,(d)], all of which display the same qualitative highest occupied molecular orbital (HOMO) splitting at the $\Gamma$ point as in Fig.~\ref{Fig:BANDS}\,(a). The calculations in Fig.~\ref{Fig:BANDS}\,(b)-(d) have an additional splitting of the HOMO bands over the entire Brillouin zone originating from the formation of antiferromagnetic order between ferromagnetic planes. The onset of antiferromagnetic order in Fig.~\ref{Fig:BANDS}\,(b)-(d) stems from the initial application of a magnetic field to split spin degeneracy despite a reduction of the applied field with subsequent calculation iterations and a convergence with no applied field. The application of a Hubbard $U$, Fig.~\ref{Fig:BANDS}\,(d), is required to fully open the gap and predict insulating behavior. The splitting observed in the $a_1$ derived bands in these calculations is allowed due to the unit cell symmetry and there being an even number of electrons per unit cell \cite{rondinelli_structure_2011}. The large cluster interaction energy indicated by the width of the $a_1$ band ($\approx 200$\,meV) clearly plays a key role in the collective transition from a triangular lattice \Shalf antiferromagnet to a system of static singlets.

The special points of the Brillouin zone used in the HT phase calculations are for a trigonal unit cell in the hexagonal setting and are $\Gamma = (000)$, $\mathrm{K} = (\frac{\bar{1}}{3} \frac{2}{3} 0)$, $\mathrm{M} = (0 \frac{1}{2} 0)$, $\mathrm{A} = (0 0 \frac{1}{2})$, $\mathrm{L} = (0 \frac{1}{2} \frac{1}{2})$, and $\mathrm{H} = (\frac{\bar{1}}{3} \frac{2}{3} \frac{1}{2})$. The special points of the Brillouin zone for a monoclinic unit cell in the \Ctm space group with unique axis $b$, cell choice 1 used in the LT calculations are $\mathrm{V} = (\frac{1}{2} 0 0)$ $\mathrm{Z} = (\frac{1}{2} \frac{1}{2} 0)$ $\Gamma = (000)$ $\mathrm{A} = (0 0 \frac{1}{2})$ $\mathrm{M} = (\frac{1}{2} \frac{1}{2} \frac{1}{2})$ $\mathrm{L} = (\frac{1}{2} 0 \frac{1}{2})$.

\FloatBarrier

\section{Spin-Peierls analysis}

\begin{figure}[h!]
	\centering
		\includegraphics[width=4in]{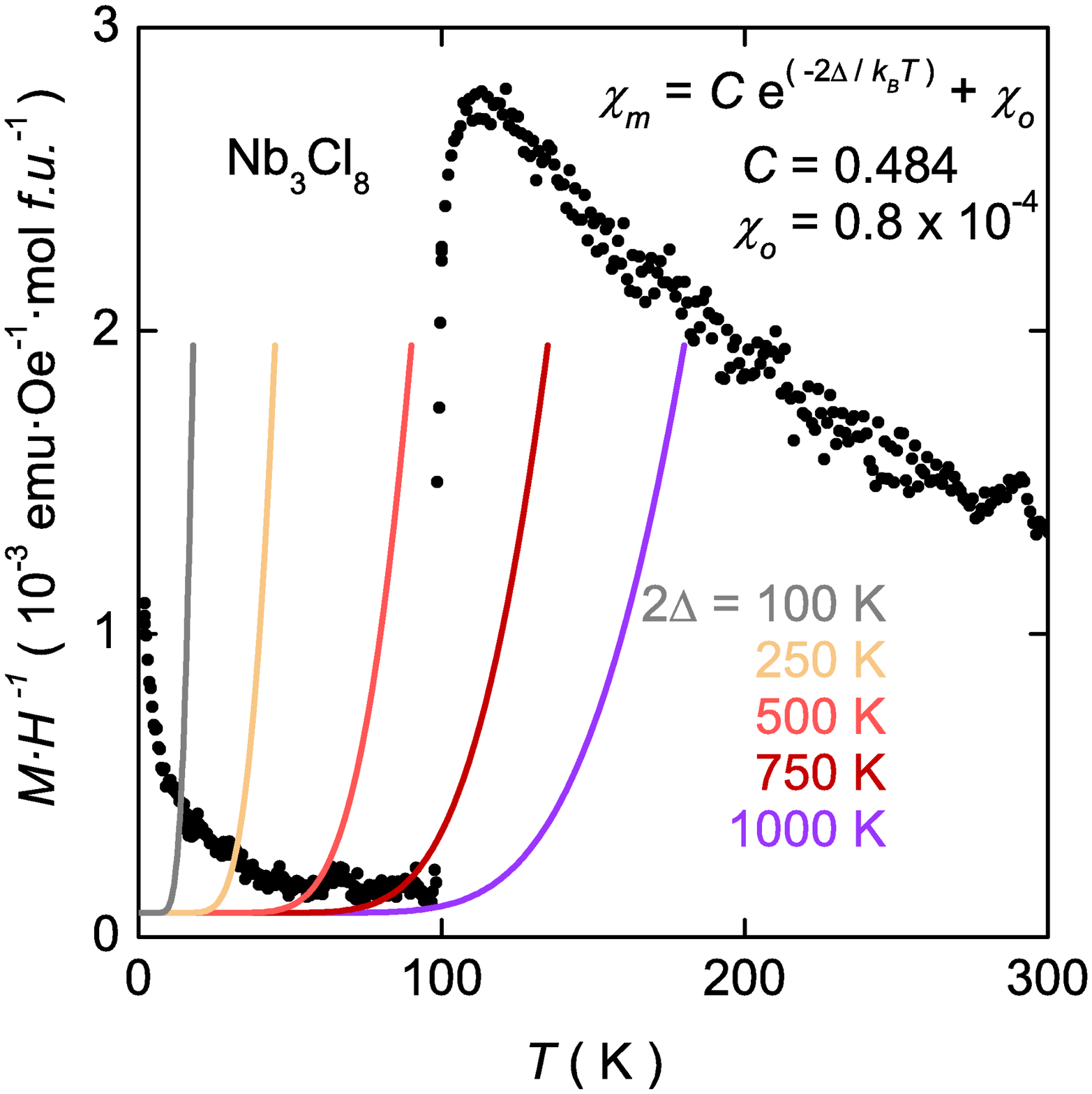}
			\caption[]{\NbC magnetic susceptibility data (black dots) as a function of temperature upon warming. The data at LT is a convolution of a Curie tail and the intrinsic behavior of the LT magnetic state of \NbC. The colored lines are calculations of the exponential behavior expected of a spin-Peierls state. As can be seen, a gap of at least $2 \Delta = 1000$\,K must exist in order to observe the magnetic behavior of a spin-Peierls state in \NbC.}
		\label{Fig:SP}
\end{figure}

The transition into a spin-Peierls (SP) state is marked by an exponential decay of the magnetic susceptibility as a function of temperature \cite{bulaevskii_spin-peierls_1978}. The SP transition temperature depends on the magnetic interaction strength and the spin excitation gap in the SP state. \NbC does not have an exponential decay into the low-temperature (LT) state---the first-order nature of the transition results in remnant domains of the HT phase upon cooling. Instead, a plot of magnetic susceptibility collected upon warming, Fig.~\ref{Fig:SP}, of the $c \parallel$~\muoH data [reproduced from main text Fig.~2\,(a)] must be used in modeling the behavior of the bulk LT state. As can be seen in Fig.~\ref{Fig:SP}, the susceptibility data is flat in temperature (with the exception of the Curie tail in $T \leq 50$\,K) up to the transition at \Tapp100\,K. To test that the magnetic susceptibility from a SP state ($\chi_{\mathrm{SP}}$) is varying minimally with temperature, a Curie-Weiss fit to the LT data up to \Teq50\,K and \Teq100\,K was performed to extract a value of $\chi_\mathrm{o} \approx \chi_{\mathrm{SP}}$ for the two temperature regions, a deconvolution of the Curie tail contribution. Values of $\chi_\mathrm{o} = 0.8$ and $0.7 \times 10^{-4}\ \mathrm{emu \cdot Oe^{-1} \cdot {mol}}\ f.u.^{-1}$ are extracted for fits up to \Teq100\,K and \Teq50\,K respectively. 

The magnetic behavior of a SP state can be approximated by the exponential \cite{baker_muon-spin_2007} $$\chi_m = C \ e^{\frac{-2 \Delta}{k_BT}} + \chi_\mathrm{o}$$
where $\chi_m$ is the molar susceptibility, $C$ is the Curie constant, $2 \Delta$ is the gap to the first excited state, $k_B$ is Boltzmann's constant, and $\chi_\mathrm{o}$ is the temperature independent contribution to the susceptibility. It is important to note that the transition temperature between HT and LT states in \NbC is not necessarily the spin-Peierls transition---since the transition in \NbC is first-order, the HT and LT magnetic behaviors need not be related in any way. Assuming the LT state in \NbC is a SP state, then calculations for various magnitudes of the spin gap show that the gap must be at least $2 \Delta = 1000$\,K in order reproduce the observed behavior. These calculations assume the Curie constant is the value extracted from the HT fit for a \Shalf system, $C = 0.484\ \mathrm{emu \cdot K \cdot mol\ \mathit{f.u.}^{-1} \cdot Oe^{-1}}$, and the temperature independent contribution is $\chi_\mathrm{o} = 0.8 \times 10^{-4}\ \mathrm{emu \cdot Oe^{-1} \cdot {mol}}\ f.u.^{-1}$.

The magnitude of the spin gap may also be estimated, assuming LT \NbC is a spin-Peierls state and the transition temperature $T_c = 100$\,K, by examining the ratio of the susceptibility above and below the transition as is done in other SP materials such as CuGeO$_3$ \cite{hase_observation_1993}. Approximating $\chi_{\mathrm{HT}} / \chi_{\mathrm{LT}} \approx R$, and rearranging $R = e^{\frac{-\Delta}{T_c}}$ to $\Delta = T_c \cdot ln(R)$ with $R \approx 30$ gives $\Delta = 100\,\mathrm{K} \cdot ln(30) = 340$\,K. This is on the same order as to what may be expected of a SP system, where the gap $\Delta \approx 1.5 \cdot T_c$.

We can also make an approximation of the exchange constant $J$ in the \NbC spin-Peierls system from the spin-phonon coupling constant given by Cross and Fisher \cite{cross_new_1979}. Here, $T_c = 0.8 \cdot J \cdot \eta$, or $J = \frac{T_c \cdot 1.25}{\eta}$. Assuming the value of $\eta$ used for CuGeO$_3$, $\eta = 0.2$, is used for \NbC, the calculation yields a $J = \frac{100\,\mathrm{K} \cdot 1.25}{0.2} = 625$\,K--- much larger than the value extracted from the Curie-Weiss fit to the HT susceptibility data of $\theta = 51.2$\,K. This discrepancy may be indicative of a strong enhancement of the exchange constants upon entering the LT state from the structural rearrangement, or that a spin-Peierls model fails to adequately explain the magnetic behavior of \NbC.

\FloatBarrier
\newpage

\bibliography{Nb3Cl8} 
\bibliographystyle{rsc} 

\end{document}